\documentclass[twocolumn, secnumarabic,amssymb, nobibnotes, aps, prd]{revtex4-2}
\usepackage{hyperref}
\usepackage{amsmath}
\usepackage{graphicx}
\usepackage{float} 
\usepackage{placeins}
\usepackage{booktabs}

\begin{document}
	
	\title{Symmetry--Protected Momentum Exchange between Dark Matter and Dark Energy}
	
	\author{Mohid Farhan}
	\email{mohidf35@gmail.com} 
	\affiliation{
		Department of Space Science, Institute of Space Technology, Islamabad, Pakistan
	}
	
\date{May 19, 2026}
	
	\begin{abstract}
		We present a particle--physics motivated realization of interacting dark energy in which a radiatively stable dark energy sector couples to weakly interacting massive particle dark matter through pure momentum exchange. The dark energy field arises as a pseudo--Nambu--Goldstone boson from a complex scalar singlet charged under a softly broken global $U(1)_S$, while dark matter is identified with an inert scalar doublet stabilized by a discrete $Z_4$ symmetry. This symmetry structure allows renormalizable dark matter--dark energy portal operators; however, requiring the dark energy field to emerge as a radiatively stable pseudo--Nambu--Goldstone boson necessitates their absence, leaving derivative interactions as the leading coupling. As a result, energy transfer between the dark sectors is absent at the background level, while momentum exchange modifies the evolution of cosmological perturbations. We implement the resulting interacting dark energy model self--consistently in the Boltzmann code \texttt{CLASS} and study its impact on the growth of structure. We find that, despite sizeable momentum exchange, the suppression of the clustering amplitude $\sigma_8$ saturates above the level required to fully resolve current low--redshift tensions. Our results demonstrate that symmetry--protected, momentum--exchange--only dark sector interactions possess an intrinsic limit on structure suppression, providing a theoretically controlled benchmark for interacting dark energy scenarios.
	\end{abstract}

	\maketitle
	
	\section{Introduction}
	
	The nature of dark matter (DM) and dark energy (DE) remains one of the central open problems in modern cosmology. While the $\Lambda$CDM paradigm successfully describes a wide range of cosmological observations, persistent tensions between early- and late-time measurements (most notably in the clustering amplitude $\sigma_8$ and the Hubble parameter $H_0$) have motivated the exploration of physics beyond a minimally coupled cosmological constant \cite{Planck2018, Riess2022, DESI2024}. Among the proposed extensions, interacting dark energy (IDE) models provide a theoretically appealing framework in which non-gravitational interactions within the dark sector modify the growth of cosmic structure while leaving the visible sector largely unaffected.
	
	A broad class of IDE scenarios introduces an explicit coupling between DM and DE through phenomenological energy transfer terms at the level of the cosmological continuity equations \cite{Amendola2000, Wetterich1995}. While such models can alleviate certain observational tensions, they often lack a well-defined particle physics origin and may suffer from instabilities, strong coupling issues, radiative sensitivity or fine-tuning\cite{Valiviita2008, He2011}. These shortcomings have motivated the study of IDE models in which energy transfer vanishes at the background level, leaving only momentum exchange between the dark sectors \cite{Simpson2010, Pourtsidou2013, Skordis2015}. Momentum-exchange IDE models are known to affect cosmological perturbations without altering the background expansion history, making them particularly attractive from an observational consistency standpoint.
	
	Despite their phenomenological appeal, most momentum-exchange IDE models are formulated at the level of effective fluids or phenomenological drag terms, with limited connection to an underlying ultraviolet-complete theory. Embedding such interactions into a consistent particle physics framework poses significant challenges, especially when dark energy is dynamical. In particular, any viable DE scalar field must remain radiatively stable against large quantum corrections, which generically drive its mass far above the Hubble scale unless protected by a symmetry \cite{Carroll1998, Kaloper2016}.
	
	A well-known mechanism to stabilize ultralight scalar fields is to realize dark energy as a pseudo--Nambu--Goldstone boson (pNGB) associated with the spontaneous breaking of a global symmetry \cite{Frieman1995, Choi2000,Lee2020}. In such constructions, the shift symmetry of the pNGB protects the scalar mass, while soft symmetry breaking generates a naturally small potential. Separately, inert scalar extensions of the Standard Model provide a minimal and well-studied framework for weakly interacting massive particle (WIMP) dark matter, stabilized by a discrete symmetry and compatible with collider, astrophysical, and cosmological constraints \cite{Deshpande1977, Barbieri2006, Gustafsson2007}.
	
	In this work, we combine these two ideas into a unified, symmetry-protected dark sector in which a pNGB dark energy field interacts with inert scalar dark matter exclusively through momentum exchange. The dark energy sector arises from a complex scalar singlet charged under a global $U(1)_S$ symmetry that is spontaneously broken at a high scale and softly broken by a dimension-two operator, ensuring radiative stability of the pNGB mass. Dark matter is identified with an inert scalar doublet stabilized by a discrete $Z_4$ symmetry. Crucially, while the imposed symmetry structure allows renormalizable dark matter--dark energy portal operators, the requirement that dark energy emerge as a radiatively stable pseudo–Nambu–Goldstone boson necessitates their absence, leaving derivative interactions as the leading coupling. As a consequence, energy transfer between dark matter and dark energy vanishes at the background level, while momentum exchange modifies the evolution of cosmological perturbations.
	
	We implement this interacting dark energy model self-consistently in the Boltzmann code \texttt{CLASS} and study its impact on the growth of large-scale structure. Our analysis demonstrates that, even for sizeable momentum-exchange rates, the suppression of $\sigma_8$ saturates above the level required to fully resolve current low-redshift tensions. This result indicates that symmetry-protected, momentum-exchange--only dark sector interactions possess an intrinsic limit on their ability to suppress structure growth.
	
	The purpose of this paper is twofold. First, we provide an explicit particle physics realization of momentum-exchange interacting dark energy that is radiatively stable and symmetry controlled. Second, we use this framework to assess the extent to which such interactions can impact cosmological observables, thereby clarifying the capabilities and limitations of this class of models. Our results establish a theoretically well-motivated benchmark for interacting dark energy scenarios and provide guidance for future extensions involving additional interactions or symmetry breaking effects.

\section{Theoretical Framework}

\subsection*{Field Content and Symmetries}

We consider an extension of the Standard Model (SM) featuring an inert scalar doublet and a complex scalar singlet, which together constitute the dark sector~\cite{Deshpande1977, Barbieri2006, Gustafsson2007}. The field content is
\begin{equation}
	H = \begin{pmatrix}
		H_1^{+} \\
		\frac{H_1^0 + i A_1^0}{\sqrt{2}}
	\end{pmatrix}, H_2 = \begin{pmatrix}
		H_2^{+} \\
		\frac{H_2^0 + i A_2^0}{\sqrt{2}}
	\end{pmatrix}, S = \begin{pmatrix}
		\frac{v_s + \rho + i \phi}{\sqrt{2}}
	\end{pmatrix}.
\end{equation}
where $H$ is the SM Higgs doublet, $H_2$ is an inert scalar doublet identified with dark matter, and $S$ is a complex scalar singlet responsible for dark energy~\cite{Belanger2022}.

The theory is endowed with a discrete $Z_4$ symmetry and a global $U(1)_S$ symmetry under which the fields transform as
\begin{equation}
	H \rightarrow H, \qquad
	H_2 \rightarrow - H_2, \qquad
	S \rightarrow i S, \qquad
	S \rightarrow e^{i\alpha} S.
\end{equation}
The $Z_4$ symmetry stabilizes the inert scalar doublet and ensures the absence of dark matter decay into Standard Model states~\cite{Belanger2008}. However, the $Z_4 \times U(1)_S$ symmetry alone does not forbid all renormalizable interactions between the inert doublet and the singlet sector. In particular, operators of the form $|S|^2 |H_2|^2$ are invariant under both symmetries and would generically be allowed~\cite{Kim2025}.

The absence of such operators in the present framework is enforced by the requirement that the dark energy field emerge as a pseudo-Nambu-Goldstone boson with a radiatively stable, ultralight mass~\cite{Frieman1995, Choi2000}. This necessitates that the global $U(1)_S$ symmetry be spontaneously broken by the singlet vacuum expectation value while being only \emph{softly} broken in the potential~\cite{tHooft1980, Giudice2008}. The soft breaking is achieved through a single dimension-two operator proportional to $\mu_{sb}^2$, which is the unique source of explicit symmetry violation in the singlet sector~\cite{Snowmass2022}. Consequently, any interaction that would generate an unsuppressed mass for the angular mode $\psi$ must be set to zero in order to preserve the pNGB nature of dark energy. Their vanishing protects the technical naturalness of the pNGB mass scale, though as we shall see, this is a phenomenological requirement rather than a symmetry-enforced selection~\cite{Carroll1998, Kaloper2016}.

Expanding the singlet field around its vacuum expectation value,
\begin{equation}
	\langle S \rangle = \frac{v_s}{\sqrt{2}}, \qquad S = \frac{1}{\sqrt{2}} \left( v_s + \rho + i \phi \right),
\end{equation}
the angular mode $\phi$ transforms nonlinearly under $U(1)_S$ and is identified with the dark energy field~\cite{Peebles1988}. At tree level, its mass is controlled entirely by the soft-breaking parameter,
\begin{equation}
	m_\phi^2 = 2 |\mu_{sb}^2|,
\end{equation}
while the radial mode acquires a mass $m_\rho^2 = \lambda_S v_s^2 + 2|\mu_{sb}^2| \approx \lambda_S v_s^2$. In the limit $\mu_{sb}^2 \rightarrow 0$, the $U(1)_S$ symmetry is restored and the dark energy field becomes exactly massless. The smallness of $m_\phi \sim H_0$ is therefore technically natural: it is protected from large radiative corrections by the fact that $\mu_{sb}^2$ is the only explicit breaking parameter, and any correction to $m_\phi^2$ must be proportional to $\mu_{sb}^2$ itself~\cite{tHooft1980}.

\subsection*{Scalar Potential and Symmetry Structure}

The most general renormalizable scalar potential consistent with gauge symmetry and the imposed $Z_4 \times U(1)_S$ charges contains all allowed operators:
\begin{align}
	V(H, H_2, S) &= \mu_1^2 |H|^2 + \mu_2^2 |H_2|^2 + \mu_S^2 |S|^2 \nonumber \\
	&\quad + \frac{\lambda_1}{2} |H|^4 + \frac{\lambda_2}{2} |H_2|^4 + \frac{\lambda_{S}}{2} |S|^4 \nonumber \\
	&\quad + \lambda_3 |H|^2|H_2|^2 + \lambda_4 |H^\dagger H_2|^2 
	+ \left[\frac{\lambda_5}{2}(H^\dagger H_2)^2 + \text{h.c.}\right] \nonumber \\
	&\quad + \lambda_{S1} |S|^2 |H|^2 + \lambda_{S2} |S|^2 |H_2|^2 \nonumber \\
	&\quad + \left[ \lambda_{S12} S^2 (H^\dagger H_2) + \lambda_{S21} S^{*2} (H^\dagger H_2) + \text{h.c.} \right] \nonumber \\
	&\quad + \left[ \lambda_{S4} S^4 + \text{h.c.} \right] + \left[ \frac{\mu_{sb}^2}{2} S^2 + \text{h.c.} \right],
	\label{eq:full_z4idsm_potential}
\end{align}
where ``h.c.'' denotes the Hermitian conjugate of the preceding terms. All portal couplings ($\lambda_{S1}$, $\lambda_{S2}$, $\lambda_{S12}$, $\lambda_{S21}$, $\lambda_{S4}$) are invariant under $Z_4$ and $U(1)_S$, and $\mu_{sb}^2$ parametrizes the soft breaking.

However, the portal operators $|S|^2|H_2|^2$ and $|S|^2|H|^2$ are incompatible with the pseudo-Nambu-Goldstone nature of dark energy. When $S$ acquires its vacuum expectation value, these terms contribute to the Coleman-Weinberg potential for the angular mode $\phi$~\cite{Coleman1973}, generating a mass-squared correction at one-loop order. The dominant contribution arises from the dark matter loop, giving
\begin{equation}
	\delta m_\phi^2 = \frac{\lambda_{S2}^2}{16\pi^2} m_{H_2}^2 \left[\ln\left(\frac{\Lambda^2}{m_{H_2}^2}\right) + \mathcal{O}(1)\right],
\end{equation}
where $\Lambda$ denotes the ultraviolet cutoff scale. Renormalization group analysis of the IDSM reveals that the model loses perturbativity at $\Lambda \sim 10^8$~GeV due to the running of the inert doublet couplings. Requiring that the quantum correction not exceed the tree-level mass, $|\delta m_\phi^2| \leq m_\phi^2 = 2|\mu_{sb}^2|$, imposes the constraint
\begin{equation}
	\lambda_{S2}^2 \leq \frac{16\pi^2 m_\phi^2}{m_{H_2}^2 \ln(\Lambda^2/m_{H_2}^2)}.
\end{equation}
For a weak-scale dark matter particle with $m_{H_2} \sim 10^2$~GeV and dark energy mass $m_\phi \sim H_0 \sim 10^{-33}$~eV $\sim 10^{-42}$~GeV, taking $\Lambda \sim 10^8$~GeV yields $\ln(\Lambda^2/m_{H_2}^2) \sim 28$ and
\begin{equation}
	|\lambda_{S2}| \lesssim \frac{4\pi \times 10^{-42}~\text{GeV}}{(10^2~\text{GeV}) \times \sqrt{28}} \sim 10^{-44}.
\end{equation}
The operators $S^2(H^\dagger H_2)$ and $S^{*2}(H^\dagger H_2)$ similarly contribute to the $\phi$ mass through mixing with the radial mode $\rho$ at comparable levels, while $S^4$ contributes to the Coleman-Weinberg potential for $\rho$ and indirectly to $\phi$ after symmetry breaking. All these effects would destroy the hierarchy between the dark energy scale and the electroweak scale~\cite{Weinberg1989}.
We therefore impose the \textbf{pNGB protection condition}: all hard portal couplings between the singlet and doublet sectors are set to zero:
\begin{equation}
	\lambda_{S1} = \lambda_{S2} = \lambda_{S12} = \lambda_{S21} = \lambda_{S4} = 0.
\end{equation}
The sole remnant of explicit breaking in the singlet sector is the soft term $\mu_{sb}^2$. This choice is motivated by naturalness: it protects the ultralight mass of the dark energy field from quantum corrections proportional to the heavy scales in the dark matter sector~\cite{tHooft1980}. However, it is \emph{not enforced by the $Z_4 \times U(1)_S$ symmetry}---the symmetry structure allows these operators; their absence should be understood as a phenomenological boundary condition necessary for a viable quintessence scenario~\cite{Carroll1998}.

The resulting potential, which we denote $V_{\rm sym}$, then separates into minimal and decoupled sectors:
\begin{align}
	V_{\text{sym}} &= \mu_1^2 |H|^2 + \mu_2^2 |H_2|^2 + \frac{\lambda_1}{2} |H|^4 + \frac{\lambda_2}{2} |H_2|^4 + \lambda_3 |H|^2 |H_2|^2 \nonumber \\
	&\quad + \lambda_4 |H^\dagger H_2|^2 + \left[ \frac{\lambda_5}{2} (H^\dagger H_2)^2 + \text{h.c.} \right] \nonumber \\
	&\quad + \mu_S^2 |S|^2 + \frac{\lambda_{S}}{2} |S|^4 + \left[ \frac{\mu_{sb}^2}{2} S^2 + \text{h.c.} \right].
	\label{eq:Vsyn}
\end{align}
where ``h.c.'' again denotes the Hermitian conjugate. We require $\mu_S^2 < 0$ to spontaneously break $U(1)_S$, giving $S$ a VEV.

\subsection*{Radiative Stability and Renormalization Group Considerations}

The radiative isolation of the dark energy sector is verified through the one-loop renormalization group equations. In the pNGB-protected limit where all singlet–doublet portal couplings vanish, the $\beta$-functions for the singlet mass parameters receive no contributions from the Higgs or inert doublet sectors:
\begin{equation}
	16\pi^2 \beta_{\mu_S^2} = 4\lambda_S \mu_S^2, \qquad 16\pi^2 \beta_{\mu_{sb}^2} = 2\lambda_S \mu_{sb}^2.
\end{equation}

The absence of terms involving $\mu_1^2$, $\mu_2^2$, $\lambda_{3,4,5}$, or gauge/Yukawa couplings confirms that the electroweak scale does not destabilize the singlet sector~\cite{Giudice2008}. This radiative isolation is a direct consequence of the imposed symmetries and the pNGB protection condition.

The complete one-loop RGEs for the model are summarized in Table~\ref{tab:RGE_one_loop} (see Appendix~\ref{app:RGE}). The decoupling of the singlet sector from SM running ensures that the smallness of $\mu_{sb}^2$ is preserved under RG evolution, with corrections proportional only to $\mu_{sb}^2$ itself~\cite{Snowmass2022}.

This framework therefore provides a consistent ultraviolet-motivated realization of momentum-exchange interacting dark energy. The absence of renormalizable energy-transfer operators---enforced by the pNGB protection condition rather than by fine-tuning---distinguishes this construction from conventional phenomenological IDE models~\cite{Amendola2000, Wetterich1995}, while the derivative nature of the leading interaction ensures background stability~\cite{Simpson2010, Pourtsidou2013, Skordis2015}.

\subsection*{Leading Derivative Interaction and Momentum Exchange}

Although renormalizable portals are absent, higher-dimensional operators consistent with the symmetries are generically expected to arise after integrating out heavy degrees of freedom~\cite{Weinberg1979}. The leading interaction between dark matter and dark energy appears at dimension six and takes the schematic form~\cite{Simpson2010, Pourtsidou2013}
\begin{equation}
	\mathcal{L}_{\rm int} =
	\frac{c_6}{\Lambda^2}
	\left( \partial_\mu |S|^2 \right)
	\left( \partial^\mu |H_2|^2 \right) ,
\end{equation}
where $\Lambda$ denotes the effective cutoff scale and $c_6$ is a dimensionless coefficient.

After spontaneous symmetry breaking and integrating out the heavy radial mode $\rho$, this operator induces derivative couplings between the pNGB field $\phi$ and the inert scalar dark matter~\cite{Skordis2015}. Importantly, such interactions do not contribute to the background energy-momentum transfer but instead generate a drag force at the level of perturbations~\cite{Simpson2010}. The resulting cosmological dynamics correspond to a pure momentum-exchange interacting dark energy scenario, as studied phenomenologically in previous works~\cite{Pourtsidou2013, Skordis2015, Linton2022}.
	
\section{Methodology}
The analysis presented in this work combines particle--physics calculations of dark matter phenomenology with cosmological evolution at the level of linear perturbations. The two aspects are treated independently and then consistently interfaced. Dark matter properties are computed using the public code \texttt{micrOMEGAs}, while the impact of dark matter--dark energy momentum exchange on cosmological observables is evaluated using a modified version of the Boltzmann solver \texttt{CLASS}. Importantly, the background cosmological evolution remains unaltered relative to $\Lambda$CDM, and all modifications enter exclusively at the level of perturbations. We also derive an expression for the DM--DE drag rate $\Gamma$.

\subsection*{Dark Matter Phenomenology using MicrOMEGAs}
The relic density $\Omega_{\rm DM} h^2$ of the dark matter candidate is computed using micrOMEGAs, a numerical package designed to evaluate dark matter observables across BSM frameworks. For each model, the particle content and interactions are specified via model files in CalcHEP format \cite{micromegas,Micromegas5}. The time evolution of the dark matter number density $n_\chi$ is governed by the Boltzmann equation, which micrOMEGAs solves numerically:
\begin{equation}
	\frac{dn_\chi}{dt} + 3Hn_\chi = -\langle \sigma v \rangle \left(n_\chi^2 - n_{\chi}^{\text{eq}\,2} \right),
\end{equation}
where $H$ is the Hubble parameter, $\langle \sigma v \rangle$ is the thermally averaged annihilation cross-section, and $n_\chi^{\text{eq}}$ is the equilibrium number density.
MicrOMEGAs computes the thermally averaged cross-section using:

\begin{equation}
	\begin{split}
		\langle \sigma v \rangle = \frac{1}{C_1}  \int_{4m_\chi^2}^{\infty} ds\, \sigma(s)
	 (s - 4m_\chi^2)\sqrt{s}\,K_1(\sqrt{s}/T),
	\end{split}
\end{equation}
where $C_1=8m_\chi^4 T K_2^2(m_\chi/T)$. $K_1$ and $K_2$ are modified Bessel functions, and $\sigma(s)$ includes all $2 \to 2$ annihilation and coannihilation channels.
The relic abundance is derived via numerical integration over temperature:
\begin{equation}
	\Omega_{\rm DM} h^2 = \frac{1.07 \times 10^9\,\text{GeV}^{-1}}{g_*^{1/2} M_{\text{Pl}} \int_{x_f}^{\infty} \frac{\langle \sigma v \rangle}{x^2} dx},
\end{equation}

where $x = m_\chi/T$, $x_f$ is the freeze-out point, $g_*$ denotes effective relativistic degrees of freedom, and $M_{\text{Pl}}$ is the Planck mass. The relic density of dark matter is constrained to be in the region: 
\begin{equation}
	\Omega_{\rm DM}h^{2}=0.1199\pm0.0027
\end{equation} 
by Planck data \cite{Planck2018}. To maintain consistency with observation, we will include these bounds in our computational analysis. 

\subsection*{Cosmological Perturbations with \texttt{CLASS}}

We implement the Z$_4$-IDSM cosmology in the \texttt{CLASS} (Cosmic Linear Anisotropy Solving System) code~\cite{Lesgourgues2011, Blas2011}, modifying the perturbation evolution to include momentum exchange between dark matter and the pNGB dark energy fluid. The implementation preserves the background evolution of standard $\Lambda$CDM while introducing drag terms in the perturbation equations \cite{Kodama1984}. We employ the Newtonian gauge throughout \cite{Ma1995}, with metric perturbations $\Psi$ and $\Phi$ defined through the line element $ds^2 = a^2(\tau)[-(1+2\Psi)d\tau^2 + (1-2\Phi)\delta_{ij}dx^i dx^j]$.

The background module requires no modification. The dimension-6 operator contributes negligibly to the homogeneous energy density, ensuring that $\rho_{\rm DM}(a)$ and $\rho_{\rm DE}(a)$ evolve independently according to the standard continuity equations~\cite{Simpson2010, Pourtsidou2013}. The dark energy equation of state is determined by the pNGB potential. Expanding the soft-breaking term in Eq.~(5) with $S=(v_s/\sqrt{2})\exp(i\phi/v_s)$ yields the effective potential
\begin{equation}
	V(\phi)=\frac{1}{2}m_\phi^2 v_s^2\sin^2\!\left(\frac{\phi}{v_s}\right),
\end{equation}
where $m_\phi^2=2|\mu_{sb}^2|$. In the small-displacement regime relevant for cosmological evolution ($\phi\ll v_s$), this reduces to the harmonic form $V(\phi)\approx\frac{1}{2}m_\phi^2\phi^2$. For $m_\phi\sim H_0$, the field is slowly rolling today with an equation of state $w(a)\simeq-0.9$; we do not impose the de Sitter limit $w=-1$ in the fluid implementation, since the $c_s^2/(1+w)$ structure in the perturbation equations becomes singular as $1+w\to 0$. The Hubble expansion $H(a)$ therefore follows the standard Friedmann equation with uncoupled dark sectors at the background level to within $|\Delta w|<0.1$. In our \texttt{CLASS} implementation, the pNGB is treated as a dynamical fluid with this background equation of state and sound speed $c_s^2=1$.

The momentum exchange enters through modified Euler equations for the velocity divergences $\theta_{\rm cdm} \equiv ik^i v_{{\rm cdm},i}$ and $\theta_{\rm fld} \equiv ik^i v_{{\rm fld},i}$. In the Newtonian gauge, the cold dark matter velocity equation becomes
\begin{equation}
	\theta_{\rm cdm}' + \mathcal{H}\theta_{\rm cdm} - k^2\Psi = a\Gamma(a)(\theta_{\rm fld} - \theta_{\rm cdm}),
	\label{eq:cdm_euler}
\end{equation}
where primes denote conformal time derivatives, $\mathcal{H} = aH$ is the conformal Hubble parameter, and $\Gamma(a)$ is the momentum-transfer rate. The dark energy fluid velocity equation acquires the opposite drag term weighted by the density ratio to ensure total momentum conservation~\cite{Skordis2015, Linton2022}:
\begin{equation}
	\theta_{\rm fld}' + \mathcal{H}(1-3c_s^2)\theta_{\rm fld} - \frac{c_s^2}{1+w}k^2\delta_{\rm fld} - k^2\Psi = aR_{\rm fld}\,\Gamma(a)(\theta_{\rm cdm} - \theta_{\rm fld}),
	\label{eq:de_euler}
\end{equation}
where $c_s^2$ is the sound speed of the pNGB (typically $c_s^2 = 1$ for a canonical scalar field), and the weight factor
\begin{equation}
	R_{\rm fld} = \frac{\rho_{\rm cdm}}{\rho_{\rm fld}(1+w)}
\end{equation}
ensures that the interaction conserves the total momentum density $\rho_{\rm cdm}\theta_{\rm cdm} + \rho_{\rm fld}(1+w)\theta_{\rm fld}$ up to Hubble drag terms.

From the dimension-6 operator $\mathcal{L}_{\rm int} = (c_6/\Lambda^2)(\partial_\mu|S|^2)(\partial^\mu|H_2|^2)$, integrating out the heavy radial mode $\rho$ yields the low-energy momentum-transfer rate
\begin{equation}
	\Gamma(a) = \frac{c_6 v_s^2}{\Lambda^2 m_{H_2}^2}\,\rho_{\rm cdm}(a)H(a) \equiv \xi\,H(a)\left[\frac{\rho_{\rm cdm}(a)}{\rho_{{\rm cdm},0}}\right],
	\label{eq:gamma_rate}
\end{equation}
where we have defined the dimensionless coupling parameter $\xi \equiv (c_6 v_s^2/\Lambda^2 m_{H_2}^2)\rho_{{\rm cdm},0}/H_0^2$. This parametrization makes explicit that $\Gamma \sim H$ when the coupling is order unity, and that the interaction strength grows with the dark matter density as expected for a contact interaction~\cite{Cruickshank2025}.

The modifications are implemented in the \texttt{perturbations.c} module of \texttt{CLASS} v3.0~\cite{Lesgourgues2011, Naeem:2026layered}. The new input parameters \texttt{c6} (dimensionless coupling constant), \texttt{vs2} (singlet VEV squared $v_s^2$ in GeV$^2$), \texttt{Lambda2} (cutoff scale squared $\Lambda^2$ in GeV$^2$), and \texttt{mH2\_2} (inert doublet mass squared $m_{H_2}^2$ in GeV$^2$) are passed through the \texttt{perturbations\_parameters\_and\_workspace} structure to the integration routines. The drag terms are added to the right-hand side of the linear perturbation system in the \texttt{perturbations\_derivs()} function, evaluated at each time step alongside the standard Einstein-Boltzmann terms.

We disable the Parameterized Post-Friedmann (PPF) formalism~\cite{Hu2007, Fang2008} by setting \texttt{use\_ppf = no} in the parameter file. The PPF scheme replaces the explicit fluid velocity $\theta_{\rm fld}$ with a single ``slipping'' variable $\Gamma_{\rm fld}$, which is incompatible with our direct coupling between $\theta_{\rm fld}$ and $\theta_{\rm cdm}$ in Eq.~(\ref{eq:de_euler}). When PPF is active, \texttt{index\_pt\_theta\_fld} is not allocated, causing segmentation faults upon accessing the velocity perturbation.

The system of differential equations is evolved using the NDF15 integrator~\cite{Blas2011}, an implicit solver optimized for stiff equations that exhibits good performance even in the tight-coupling regime $\Gamma \gtrsim H$. However, when $\Gamma \gg H$, the strong coupling leads to rapid oscillations in the relative velocity $(\theta_{\rm cdm} - \theta_{\rm fld})$ that exhaust the adaptive step-size controller, resulting in ``step size too small'' errors. This defines the practical upper limit of the effective field theory. We find that stable integration requires $\xi \lesssim \mathcal{O}(10)$, corresponding to $\Gamma \lesssim 10H$ at $z \sim 0$.

The matter power spectrum $P(k)$ and CMB angular power spectra $C_\ell^{TT,TE,EE}$ are extracted using the standard \texttt{CLASS} output routines. The linear growth factor $f(z) \equiv d\ln D/d\ln a$ and the $\sigma_8$ normalization are computed from the matter power spectrum at $z=0$ following the standard definitions~\cite{Planck2018}.

\subsection*{Phenomenological Treatment of the DM--DE Drag Rate}

The momentum-transfer rate $\Gamma(a)$ entering the cosmological perturbation equations originates from the dimension-6 interaction between the dark sectors. We outline the connection between the microscopic EFT and the macroscopic fluid description.

The lowest-dimensional operator capable of mediating DM--DE momentum exchange and consistent with the imposed $Z_4 \times U(1)_S$ symmetry is:
\begin{equation}
	\mathcal{L}_{\rm int} = \frac{c_6}{\Lambda^2}(\partial_\mu |S|^2)(\partial^\mu |H_2|^2),
	\label{eq:dim6_operator}
\end{equation}
where $c_6 \sim \mathcal{O}(1)$ is a Wilson coefficient and $\Lambda$ denotes the EFT cutoff scale~\cite{Weinberg1979, Simpson2010}. At energies $E \ll m_\rho$, this operator induces elastic scattering between dark matter and gradients of the pNGB field, leading to momentum exchange without background energy transfer~\cite{Simpson2010, Pourtsidou2013, Skordis2015}.

Following standard treatments of momentum-exchange IDE, we parametrize the drag force through an effective rate $\Gamma(a)$ appearing in the Euler equations~\cite{Skordis2015, Linton2022}.

Dimensional analysis suggests $\Gamma \propto H$, with proportionality constants of order unity depending on the details of the EFT matching. For computational implementation, we employ the explicit dimensionless parametrization
\begin{equation}
	\frac{\Gamma(a)}{H(a)} = \xi\left[\frac{\rho_{\rm cdm}(a)}{\rho_{\rm cdm,0}}\right].
	\label{eq:gamma_normalized_revised}
\end{equation}
In the matter-dominated era, $\rho_{\rm cdm}(a)/\rho_{\rm cdm,0} = a^{-3}$, so the drag rate scales as $\Gamma(a) \propto a^{-3}H(a)$. In our numerical exploration, we treat $\xi$ as a constant and vary it over a range corresponding to weak-to-moderate coupling, $\Gamma \lesssim \mathcal{O}(H)$. This captures the essential phenomenology of momentum-exchange IDE while maintaining perturbative control of the EFT. 

The impact of the interaction on structure formation becomes significant when $\Gamma \sim H$ at late times, where dark energy perturbations are non-negligible and the drag term competes with Hubble friction in the dark matter Euler equation. As we demonstrate below, this leads to an asymptotic limit on the suppression of $\sigma_8$ that is insufficient to resolve the $S_8$ tension.

\section{Results}
\label{sec:results}

We now present the numerical analysis of the Z$_4$-IDSM cosmology, proceeding from the determination of the dark matter mass to the implications for large-scale structure. We first identify the parameter space yielding the observed dark matter relic density using \texttt{MicrOMEGAs}~\cite{micromegas}, then verify the radiative stability of the dark energy sector through renormalization group evolution, and finally implement the full cosmology in \texttt{CLASS} to assess the impact on the clustering amplitude $\sigma_8$.

\subsection*{Viable Dark Matter Masses}

The inert doublet mass $m_{H_2}$ and the mass splitting between the neutral CP-even ($H_2^0$) and CP-odd ($A_2^0$) components are constrained by the requirement of obtaining the observed dark matter relic density, $\Omega_{\rm DM}h^2 = 0.1199 \pm 0.0027$~\cite{Planck2018}, within the Z$_4$-IDSM framework. We parameterize the mass degeneracy through $\delta \equiv m_{A_2^0} - m_{H_2^0}$. In our model, the CP-even candidate is the dark matter candidate and therefore is less massive than its CP-odd counterpart, so $\delta \geq 0$.

\begin{figure}[H]
	\includegraphics[width=\columnwidth]{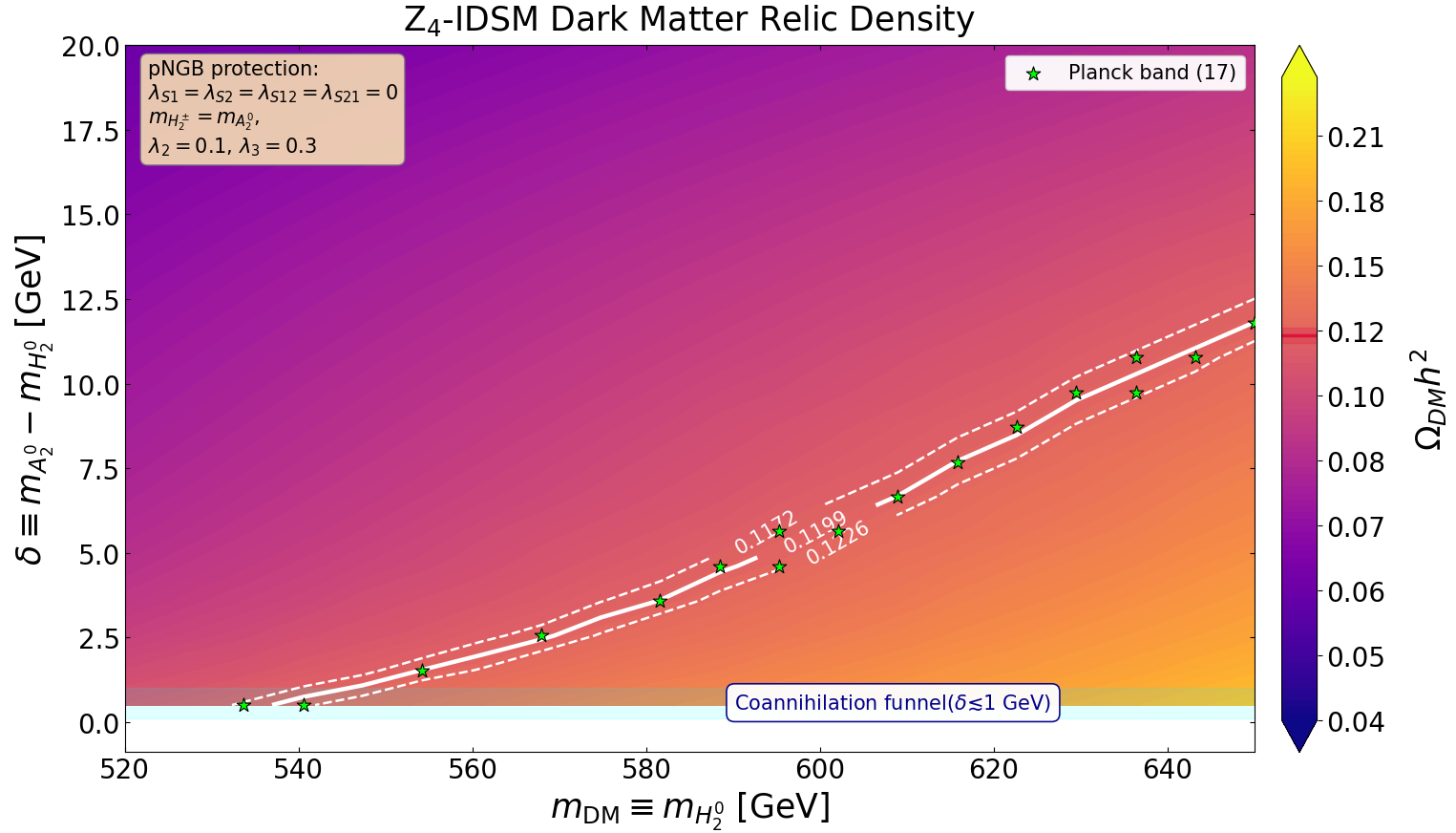}
	\caption{Dark matter relic density $\Omega_{\rm DM}h^2$ in the Z$_4$-IDSM as a function of $m_{\rm DM} \equiv m_{H_2^0}$ and mass splitting $\delta$, with pNGB protection conditions $\lambda_{S1} = \lambda_{S2} = \lambda_{S12} = \lambda_{S21} = 0$ enforced. The horizontal band indicates the coannihilation funnel ($\delta \lesssim 1$ GeV) where efficient coannihilation between $H_2^0$, $A_2^0$, and $H_2^\pm$ brings the relic density into the Planck band (green stars, $\Omega_{\rm DM}h^2 = 0.1199 \pm 0.0027$). The white contour lines indicate constant relic density values.}
	\label{fig:relic_2d}
\end{figure}

Figure~\ref{fig:relic_2d} shows the relic density in the $m_{\rm DM}$-$\delta$ plane, computed using \texttt{micrOMEGAs} with the pNGB protection conditions ($\lambda_{S1} = \lambda_{S2} = \lambda_{S12} = \lambda_{S21} = 0$) enforced. For small mass splittings ($\delta \lesssim 1$ GeV, the "coannihilation funnel"), coannihilation between the inert doublet states efficiently depletes the thermal abundance, allowing agreement with Planck constraints within the perturbative regime. For larger splittings ($\delta \gtrsim 1$ GeV), the relic density increases with $m_{\rm DM}$, requiring careful selection of the Higgs portal coupling $\lambda_L = \lambda_3 + \lambda_4 + \lambda_5$ to maintain perturbativity.

\begin{figure}[H]
	\includegraphics[width=\columnwidth]{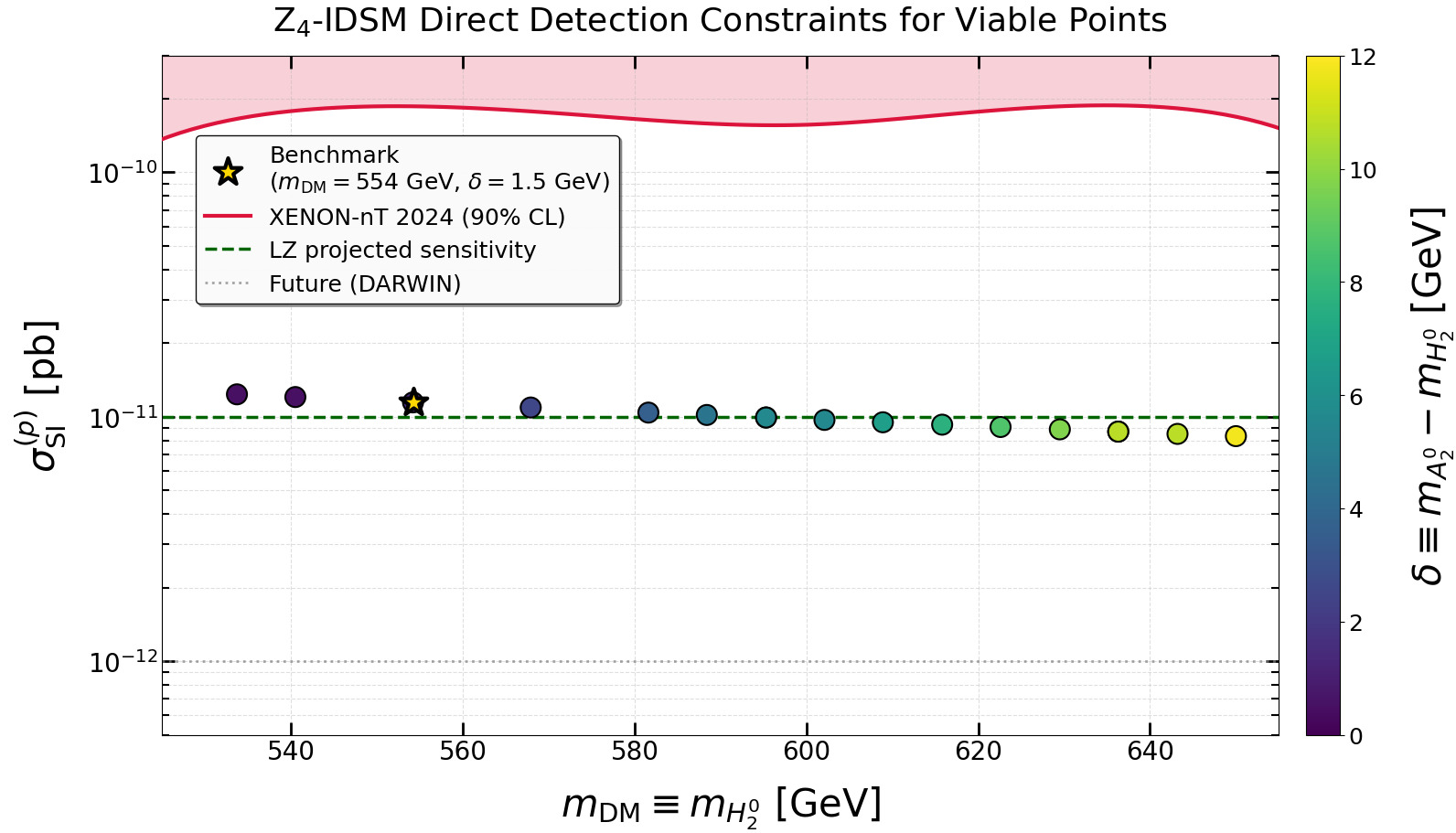}
	\caption{Direct detection constraints for Z$_4$-IDSM viable points. The colored points indicate parameter combinations satisfying $\Omega_{\rm DM}h^2 = 0.1199 \pm 0.0027$, with color coding showing the mass splitting $\delta$. The red curve shows an interpolated approximation to the published XENONnT 2024 90\% CL exclusion contour, constructed from sampled benchmark points using the corresponding direct-detection exclusion thresholds in micromegas. The green dashed line indicates the projected LZ sensitivity. The benchmark star ($m_{\rm DM} = 554$ GeV, $\delta = 1.5$ GeV) lies more than one order of magnitude below the current exclusion limit.}
	\label{fig:direct_detection}
\end{figure}

Having established the relic density constraint, we must verify compatibility with direct detection limits. For $m_{\rm DM} \sim 550$ GeV with non-zero Higgs portal coupling, the spin-independent DM-proton cross-section is subject to current bounds from LZ and XENONnT. Figure~\ref{fig:direct_detection} shows the spin-independent cross-section $\sigma_{\rm SI}^{(p)}$ for viable parameter points satisfying the relic density constraint, compared against the XENONnT 2024 90\% exclusion limit.

Table~\ref{tab:viable_points} lists representative viable candidates satisfying both the relic density constraint and direct detection bounds. All points yield $\sigma_{\rm SI} \sim 10^{-11}$ pb, comfortably below the XENONnT 2024 limit of $\sim 10^{-10}$ pb for this mass range.

\begin{table}[htbp]
	\centering
	\caption{Viable Z$_4$-IDSM candidates satisfying relic density and direct detection constraints. All spin-independent proton cross-sections are below the XENONnT 2024 90\% CL exclusion limit. For Higgs-portal interactions, $\sigma_{\rm SI}^{(p)} \approx \sigma_{\rm SI}^{(\rm nucleon)}$ (isoscalar coupling), so the quoted values correspond directly to the experimental nucleon bounds.}
	\label{tab:viable_points}
	\begin{tabular}{cccc}
		\hline
		$m_{\rm DM}$ [GeV] & $\delta$ [GeV] & $\Omega_{\rm DM}h^2$ & $\sigma_{\rm SI}^{(p)}$ [pb] \\
		\hline
		533.7 & 0.50 & 0.118 & $1.23 \times 10^{-11}$ \\
		540.5 & 0.50 & 0.122 & $1.20 \times 10^{-11}$ \\
		554.2 & 1.53 & 0.120 & $1.14 \times 10^{-11}$ \\
		567.9 & 2.55 & 0.119 & $1.09 \times 10^{-11}$ \\
		581.6 & 3.58 & 0.120 & $1.04 \times 10^{-11}$ \\
		588.4 & 4.61 & 0.119 & $1.02 \times 10^{-11}$ \\
		595.3 & 4.61 & 0.122 & $9.92 \times 10^{-12}$ \\
		595.3 & 5.63 & 0.118 & $9.92 \times 10^{-12}$ \\
		602.1 & 5.63 & 0.121 & $9.70 \times 10^{-12}$ \\
		608.9 & 6.66 & 0.120 & $9.48 \times 10^{-12}$ \\
		615.8 & 7.68 & 0.120 & $9.27 \times 10^{-12}$ \\
		622.6 & 8.71 & 0.119 & $9.07 \times 10^{-12}$ \\
		629.5 & 9.74 & 0.119 & $8.87 \times 10^{-12}$ \\
		636.3 & 9.74 & 0.122 & $8.69 \times 10^{-12}$ \\
		636.3 & 10.76 & 0.118 & $8.69 \times 10^{-12}$ \\
		643.2 & 10.76 & 0.121 & $8.50 \times 10^{-12}$ \\
		650.0 & 11.79 & 0.120 & $8.32 \times 10^{-12}$ \\
		\hline
	\end{tabular}
\end{table}

We note that for all 17 benchmark points, micrOMEGAs prompted ``Not excluded by DD experiments  at 90\% level '', confirming the experimental viability of these points. Combining the constraints from relic density, perturbativity, and direct detection, we select the benchmark point
\begin{equation}
	m_{\rm DM} = 554 \text{ GeV}, \quad \delta = 1.5 \text{ GeV}
	\label{eq:dm_params}
\end{equation}
since this point yields the relic density closest to $\Omega_{\rm DM}h^2 = 0.1199$.

\subsection*{Radiative Stability of Dark Energy}

The radiative isolation of the dark energy sector is verified by evolving the mass parameters $\mu_S^2$ and $\mu_{sb}^2$ from the electroweak scale to the cutoff $\Lambda \sim 10^8$~GeV where the IDSM loses perturbativity. We used the mathematica package called \texttt{SARAH} to compute the Renormalized Group Equations (RGEs). The one-loop RGEs receive no contributions from the Higgs or inert doublet sectors:
\begin{equation}
	16\pi^2 \beta_{\mu_S^2} = 4\lambda_S \mu_S^2, \qquad 16\pi^2 \beta_{\mu_{sb}^2} = 2\lambda_S \mu_{sb}^2.
	\label{eq:singlet_rges}
\end{equation}

Table~\ref{tab:rge_running} demonstrates this evolution for representative initial conditions with $\lambda_S = 0.5$. The hierarchy $|\mu_S^2| \sim v_s^2 \gg |\mu_{sb}^2| \sim H_0^2$ is preserved with less than $20\%$ variation over the entire perturbative regime, confirming that the ultralight dark energy scale remains technically natural and is not destabilized by quantum corrections from the electroweak sector. The key feature is the absence of terms involving $\mu_1^2$, $\mu_2^2$, $\lambda_{3,4,5}$, or gauge/Yukawa couplings in Eqs.~(\ref{eq:singlet_rges}), ensuring the singlet sector evolves independently of the SM and dark matter parameters~\cite{Giudice2008, Snowmass2022}.

\begin{table}
	\caption{Renormalization group evolution of singlet mass parameters from the electroweak scale to $\Lambda = 10^8$~GeV for $\lambda_S = 0.5$. The hierarchy $|\mu_S^2|/|\mu_{sb}^2| \sim 10^{101}$ is preserved throughout.}
	\label{tab:rge_running}
	\begin{tabular}{lccc}
		\hline\noalign{\smallskip}
		Scale $\mu$ & $|\mu_S^2|$ [GeV$^2$] & $|\mu_{sb}^2|$ [GeV$^2$] & $|\mu_S^2/\mu_{sb}^2|$ \\
		\noalign{\smallskip}\hline\noalign{\smallskip}
		$m_Z = 91$~GeV & $2.5 \times 10^{17}$ & $1.0 \times 10^{-84}$ & $2.5 \times 10^{101}$ \\
		$10^4$~GeV & $2.6 \times 10^{17}$ & $1.0 \times 10^{-84}$ & $2.6 \times 10^{101}$ \\
		$10^8$~GeV & $3.0 \times 10^{17}$ & $1.2 \times 10^{-84}$ & $2.5 \times 10^{101}$ \\
		\noalign{\smallskip}\hline
	\end{tabular}
\end{table}

\subsection*{Structure Formation and the $S_{8}$ Tension}

\begin{figure}[t]
	\includegraphics[width=\columnwidth]{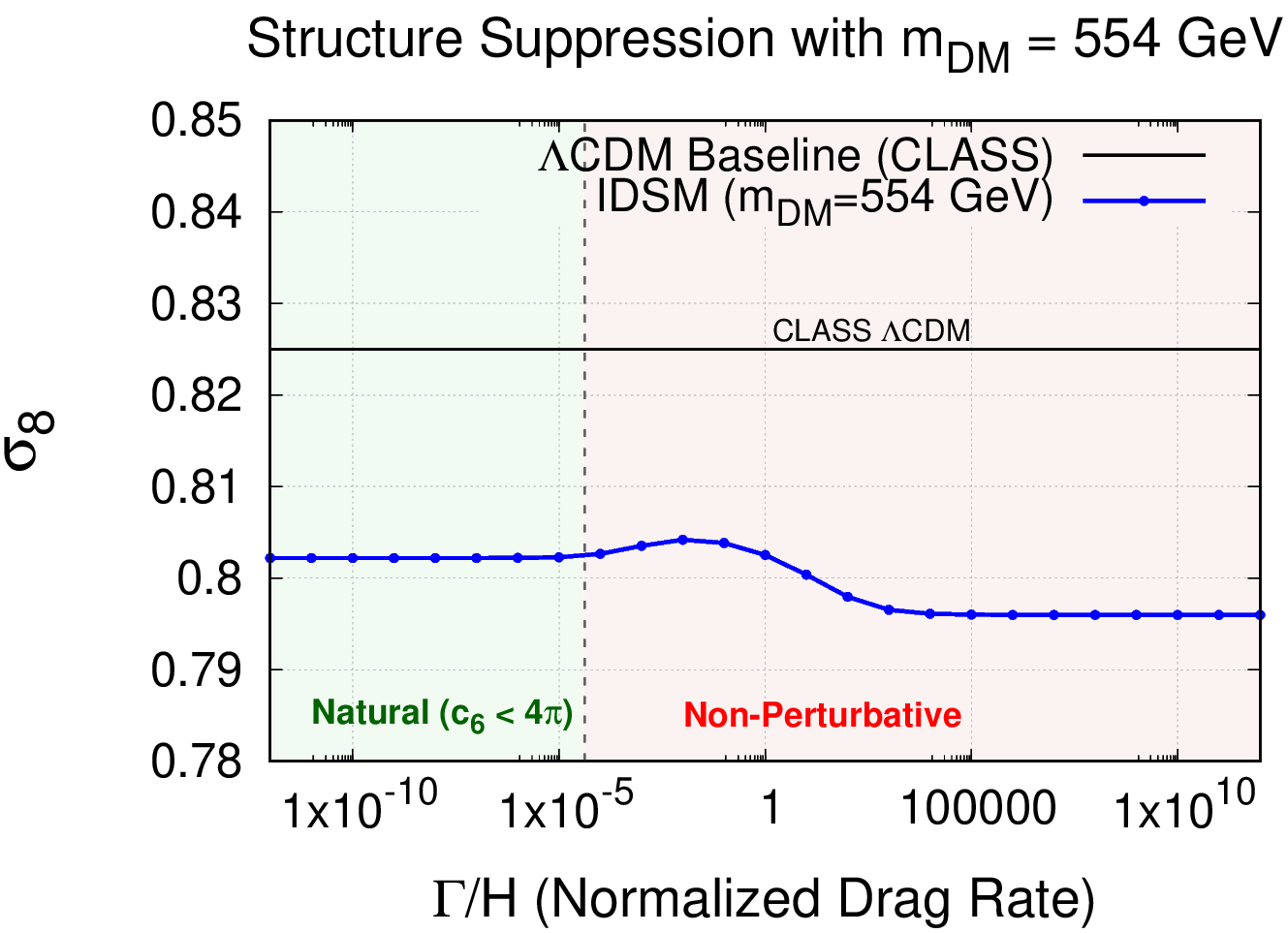}
	\caption{Clustering amplitude $\sigma_8$ as a function of the normalized drag rate $\Gamma/H$ for the Z$_4$-IDSM with benchmark parameters $m_{\rm DM} = 554$~GeV and $\delta = 1.5$~GeV from Eq.~(\ref{eq:dm_params}). The horizontal black line shows the \texttt{CLASS} $\Lambda$CDM baseline $\sigma_8^{\Lambda\text{CDM}} = 0.825$. The green (red) shaded region indicates perturbative (non-perturbative) values of the EFT coupling $c_6$. The IDSM prediction saturates at $\sigma_8 \approx 0.795$ for $\Gamma/H \gtrsim 10$.}
	\label{fig:sigma8}
\end{figure}

We implement the momentum-exchange cosmology in \texttt{CLASS} using the cosmological parameters:

$$(h, \Omega_b h^2, \Omega_c h^2, n_s, A_s) = (0.67810, 0.0224, 0.12, 0.966, 2.1\times10^{-9}),$$

yielding $\sigma_8^{\Lambda\text{CDM}} = 0.825$ and $\Omega_m = 0.3$, with benchmark DM parameters from Eq.~(\ref{eq:dm_params}) ($m_{\rm DM} = 554$~GeV, $\delta = 1.5$~GeV). Figure~\ref{fig:sigma8} shows the resulting clustering amplitude as a function of the dimensionless drag rate $\Gamma/H$, where we have absorbed the microphysics into the effective parameter $\xi_{\rm eff}$ encoding $c_6$, $v_s$, $\Lambda$, and $m_{\rm DM}$.

The pure $\Lambda$CDM baseline computed with our \texttt{CLASS} settings yields $\sigma_8^{\Lambda\text{CDM}} = 0.825$. When dark energy is treated as a dynamical fluid with equation of state $w(a)$ and perturbations, but without momentum exchange ($\Gamma/H \to 0$), the baseline shifts to $\sigma_8 \approx 0.805$ due to the non-vanishing sound speed of the pNGB. As the drag rate increases, $\sigma_8$ initially exhibits mild enhancement at intermediate $\Gamma/H \sim 0.1-1$, which arises from transient momentum transfer from the DE sound waves ($c_s^2=1$) into the DM sector. This effect temporarily boosts the effective gravitational source term before velocity equilibration dominates, before decreasing as momentum transfer suppresses structure growth. The suppression saturates at $\sigma_8 \approx 0.795$.

The naturally achievable suppression in the perturbative regime ($c_6 \leq 4\pi$),
\begin{equation}
	\frac{\sigma_8^{\rm IDSM}}{\sigma_8^{\Lambda\text{CDM}}} \approx \frac{0.804}{0.825} \approx 0.974,
\end{equation}
corresponds to a $\sim 2.6\%$ reduction in clustering amplitude, while the maximum achievable suppression in the strongly coupled regime,
\begin{equation}
	\frac{\sigma_8^{\rm IDSM}}{\sigma_8^{\Lambda\text{CDM}}} \approx \frac{0.795}{0.825} \approx 0.963,
\end{equation}
yields $\sim 3.6\%$ reduction. Expressed in terms of the $S_8 \equiv \sigma_8\sqrt{\Omega_m/0.3}$ parameter preferred by weak-lensing surveys, our model predicts $S_8 \approx 0.795$ at saturation (for $\Omega_m=0.3$), compared to the Planck 2018 value $S_8 = 0.811\pm0.006$~\cite{Planck2018} 
and recent weak-lensing measurements from KiDS-1000 
($S_8 = 0.759^{+0.024}_{-0.021}$)~\cite{KiDS2021} and DES-Y3 
($S_8 = 0.776\pm0.017$)~\cite{DESY3}. While the relative suppression of $\sim 3.6\%$ is robust against variations in the absolute $\sigma_8$ normalization, this falls short of the $\sim 5-10\%$ reduction (to $S_8 \sim 0.73-0.77$) required to fully resolve the tension with late-time probes. The numerical integration yields $\sigma_8 = 0.795 \pm 0.003$ for $\xi=10$ (CLASS integration precision). Systematic variation of cosmological parameters within Planck 2018 $1\sigma$ bounds shifts this by $\pm 0.01$. The saturation reflects a fundamental limitation: once $\Gamma \gtrsim H$, the dark sectors achieve velocity equilibrium and further coupling increases do not modify the growth history.

Our results demonstrate that symmetry-protected, momentum-exchange IDE scenarios possess an intrinsic ceiling on structure suppression. While theoretically controlled and free from background instabilities, they cannot---within their domain of validity---fully accommodate low-redshift cosmological data.

\section{Conclusion}
\label{sec:conclusion}

We have constructed a symmetry-protected framework for interacting dark energy in which dark matter and dark energy couple exclusively through momentum exchange. The Z$_4$-IDSM model unifies two well-motivated extensions of the Standard Model: an inert scalar doublet stabilized by a discrete $Z_4$ symmetry provides the dark matter candidate, while a complex scalar singlet with softly broken global $U(1)_S$ yields a radiatively stable pseudo-Nambu-Goldstone boson for dark energy. This structure forbids renormalizable energy-transfer operators, ensuring background stability, while admitting a unique dimension-6 derivative interaction that generates observable effects on the growth of structure.

Our analysis reveals a fundamental tension in such constructions. The same symmetry structure that protects the ultralight dark energy scale from quantum corrections also suppresses the strength of dark sector interactions. We identified 17 feasible dark matter candidates, with the most viable being $m_{\rm DM} = 554$~GeV and demonstrated that the model achieves the observed relic density through coannihilation in the nearly-degenerate regime. Renormalization group evolution confirms that the singlet sector remains radiatively isolated from the electroweak scale, preserving the technical naturalness of the dark energy mass $m_\phi \sim H_0$.

However, when implemented in the Boltzmann code \texttt{CLASS}, the momentum-exchange interaction yields at most $\sim 3.6\%$ suppression of the clustering amplitude $\sigma_8$ relative to $\Lambda$CDM, saturating at $\sigma_8 \approx 0.795$ in the strong-coupling limit. This falls short of the $\sim 5$--$10\%$ reduction suggested by low-redshift observations. The saturation arises because momentum-exchange IDE efficiently equilibrates the dark sector velocities when $\Gamma \gtrsim H$, after which further increases in the coupling do not modify the growth history.

Several avenues remain for extending this framework. Introducing mild energy transfer through carefully tuned portal couplings could enhance structure suppression while maintaining background stability, though at the cost of reintroducing fine-tuning. Alternatively, generalizing the singlet sector to include multiple pNGBs or modifying the early-time dynamics through out-of-equilibrium effects may open new parameter space. More broadly, our results suggest that resolving the $S_8$ tension through dark sector interactions may require explicitly breaking the protective symmetries that render such models theoretically attractive, presenting a genuine dilemma for particle physics model-building.

The Z$_4$-IDSM serves as a theoretically controlled benchmark: it demonstrates that even maximally symmetry-protected IDE scenarios face intrinsic limits on their observational impact. Future work must either identify mechanisms to circumvent these limits or confront the possibility that the $S_8$ tension originates from sources beyond the dark sector.

\section{Appendix}
	
\appendix

\begin{table*}[t]
	\caption{\label{tab:RGE_one_loop} One-loop renormalization group equations for the Z$_4$-IDSM model. Yukawa couplings are shown for a single generation.}
	\begin{ruledtabular}
		\begin{tabular}{ll}
			Parameter & $\beta^{(1)} \times 16\pi^2$ \\
			\hline
			$g_1$ & $\frac{21}{10} g_1^3$ \\
			$g_2$ & $-3 g_2^3$ \\
			$g_3$ & $-7 g_3^3$ \\
			$\mu_{1}^2$ & $-\frac{9}{10}g_1^2\mu_1^2 - \frac{9}{2}g_2^2\mu_1^2 + 6\lambda_1\mu_1^2 + 4\lambda_3\mu_2^2 + 2\lambda_4\mu_2^2 + 6\mu_1^2 \text{Tr}(Y_u^2 + Y_d^2) + 2\mu_1^2 \text{Tr}(Y_e^2)$ \\
			$\mu_{2}^2$ & $4\lambda_3\mu_1^2 + 2\lambda_4\mu_1^2 - \frac{9}{10}g_1^2\mu_2^2 - \frac{9}{2}g_2^2\mu_2^2 + 6\lambda_2\mu_2^2$ \\
			$\mu_S^2$ & $4\lambda_S \mu_S^2$ \\
			$\mu_{sb}^2$ & $2\lambda_S \mu_{sb}^2$ \\
			$Y_u$ & $Y_u [ \frac{3}{2}(Y_u^2 - Y_d^2) + 3\text{Tr}(Y_u^2 + Y_d^2) + \text{Tr}(Y_e^2) - \frac{17}{20}g_1^2 - \frac{9}{4}g_2^2 - 8g_3^2 ]$ \\
			$Y_d$ & $Y_d [ \frac{3}{2}(Y_d^2 - Y_u^2) + 3\text{Tr}(Y_u^2 + Y_d^2) + \text{Tr}(Y_e^2) - \frac{1}{4}g_1^2 - \frac{9}{4}g_2^2 - 8g_3^2 ]$
		\end{tabular}
	\end{ruledtabular}
\end{table*}

\section{Renormalization Group Equations and Radiative Stability}
\label{app:RGE}

The decoupling of the singlet sector is apparent from Table~\ref{tab:RGE_one_loop}: $\beta_{\mu_S^2}$ and $\beta_{\mu_{sb}^2}$ depend only on singlet self-couplings $\lambda_S$, with no contributions from Higgs/inert doublet parameters or SM gauge/Yukawa couplings. This confirms the radiative isolation essential for the technical naturalness of the dark energy scale.

\subsection{Two-Loop Contributions}

For completeness, we note that the evolution of the singlet mass parameters including two-loop effects takes the form:
\begin{equation}
	\frac{d\mu_S^2}{dt} = \frac{1}{16\pi^2} \left[ 4\lambda_S \mu_S^2 - \frac{10\lambda_S^2 \mu_S^2}{16\pi^2} \right],
\end{equation}
\begin{equation}
	\frac{d\mu_{sb}^2}{dt} = \frac{1}{16\pi^2} \left[ 2\lambda_S \mu_{sb}^2 - \frac{6\lambda_S^2 \mu_{sb}^2}{16\pi^2} \right],
\end{equation}
where the two-loop terms are suppressed by an additional factor of $16\pi^2$ and do not affect the qualitative conclusion of radiative stability.

\subsection{Observation on Radiative Stability}

As evidenced by the derived $\beta$-functions, the evolution of $\mu_S^2$ and $\mu_{sb}^2$ is entirely self-contained within the singlet sector. The absence of terms involving the SM Higgs mass ($\mu_1^2$), electroweak gauge couplings ($g_i$), or the top Yukawa coupling ($Y_u$) confirms that the dark energy scale is technically natural. In this construction, the quintessence scale is protected from the Standard Model hierarchy problem, as no loop diagrams link the singlet mass to the ultraviolet-sensitive sectors of the SM.

\section{Symmetry Structure and Technical Naturalness}
\label{app:symmetry}

The model's technical naturalness relies on a nested symmetry structure. We define the $Z_4$ group as the set of discrete rotations $\{1, \omega, \omega^2, \omega^3\}$ where $\omega = e^{i\pi/2}$.

The fields transform under $Z_4$ according to their assigned charges $Q$:
\begin{align}
	S &\to \omega \cdot S = iS \quad &(Q=1), \\
	H_2 &\to \omega^2 \cdot H_2 = -H_2 \quad &(Q=2), \\
	\text{SM Fields} &\to \omega^0 \cdot \text{SM} = \text{SM} \quad &(Q=0).
\end{align}

The renormalizable potential is split into two parts based on their symmetry properties:
\begin{enumerate}
	\item \textbf{The $Z_4$-invariant sector:} Terms like $|S|^2$ and $|S|^4$ are invariant under the full group because the total charge sums to $0 \pmod 4$. This sector respects the global $U(1)_s$ and keeps the pNGB $\phi$ massless.
	
	\item \textbf{The soft-breaking term:} We introduce the dimension-2 operator:
	\begin{equation}
		\mathcal{L}_{\rm soft} \supset \frac{\mu_{sb}^2}{2} (S^2 + S^{\dagger 2}).
	\end{equation}
	Under a $Z_4$ transformation $S \to iS$, the term transforms as $S^2 \to -S^2$. Since the Lagrangian is not invariant under the $90^\circ$ rotation ($\omega$), the $Z_4$ symmetry is explicitly broken.
\end{enumerate}

Crucially, the $S^2$ term is invariant under the subgroup $Z_2 \subset Z_4$. If we apply the $Z_4$ transformation twice ($\omega^2 = -1$):
\begin{equation}
	S \to -S \implies S^2 \to (-S)^2 = S^2.
\end{equation}
This $180^\circ$ rotation remains a valid symmetry of the full Lagrangian. Because $H_2$ carries a $Z_4$ charge of $2$, it also transforms under this residual $Z_2$ as:
\begin{equation}
	H_2 \xrightarrow{Z_2} -H_2.
\end{equation}
This ensures that $H_2$ (the dark matter) is the lightest parity-odd particle under the surviving $Z_2$. Since all Standard Model particles are even under $Z_2$, the stability of dark matter is mathematically guaranteed despite the mass-generating break in the singlet sector.

\subsection{Technical Naturalness and Dimensional Analysis}
\label{app:naturalness}

The stability of the dark energy scale $\mu_{sb}^2 \ll v^2$ is guaranteed by the ``soft'' nature of the breaking term. In quantum field theory, the sensitivity of the vacuum to radiative corrections is governed by the dimensionality of the operators.

The term $V_{\rm soft} = \frac{\mu_{sb}^2}{2}(S^2 + S^{\dagger 2})$ is a mass-dimension 2 operator. Its contribution to the renormalization group equations of the dimensionless couplings $\lambda_i$ is identically zero at the renormalizable level:
\begin{equation}
	\frac{d \lambda_i}{d \ln \mu} \propto \mathcal{F}(\lambda_j, g_k, y_l) \neq \mathcal{G}(\mu_{sb}^2).
\end{equation}
Because dimensionless couplings cannot depend linearly on mass-squared parameters (due to dimensional consistency), the smallness of the dark energy scale does not ``pollute'' the stability of the Higgs quartic coupling or the gauge sector.

According to 't Hooft's prescription, a parameter is technically natural if setting it to zero increases the symmetry of the theory~\cite{tHooft1980}: 
\begin{itemize}
	\item If $\mu_{sb}^2 \to 0$, the potential gains an exact $U(1)_s$ and $Z_4$ symmetry.
	\item If the portal couplings $\lambda_{S1}, \lambda_{S2} \to 0$, the singlet sector becomes completely decoupled from the SM.
\end{itemize}
Thus, the tiny value of $\mu_{sb}^2$ is protected from large corrections by the fact that it is the \textit{unique} source of symmetry breaking in the singlet sector. Any radiative correction to $\mu_{sb}^2$ must be proportional to $\mu_{sb}^2$ itself, ensuring that once it is set small, it remains small at all scales~\cite{Giudice2008, Snowmass2022}.

\section{Detailed Derivation of the Singlet Sector Mass Spectrum}
\label{app:mass_spectrum}

We derive the tree-level masses for the radial mode $\rho$ and the pseudo-Nambu-Goldstone boson (pNGB) $\phi$. We consider the singlet scalar potential:
\begin{equation}
	V_S = \mu_S^2 |S|^2 + \frac{\lambda_S}{2} |S|^4 + \frac{\mu_{sb}^2}{2} (S^2 + S^{\dagger 2}),
\end{equation}
where $S$ is expanded around its vacuum expectation value $v_s$ as:
\begin{equation}
	S = \frac{v_s + \rho + i\phi}{\sqrt{2}}.
\end{equation}

\subsection{Field Expansion}

\begin{equation}
	|S|^2 = S S^\dagger = \frac{1}{2}(v_s + \rho + i\phi)(v_s + \rho - i\phi) = \frac{1}{2}\left[ (v_s + \rho)^2 + \phi^2 \right].
\end{equation}
Expanding this gives:
\begin{equation}
	|S|^2 = \frac{1}{2}v_s^2 + v_s \rho + \frac{1}{2}\rho^2 + \frac{1}{2}\phi^2.
\end{equation}

We square the result from the previous step, keeping only terms up to quadratic order in the fields (as cubic and quartic terms do not contribute to tree-level masses):
\begin{align}
	|S|^4 &= \left( \frac{1}{2}v_s^2 + v_s \rho + \frac{1}{2}\rho^2 + \frac{1}{2}\phi^2 \right)^2 \nonumber \\
	&\approx \frac{1}{4}v_s^4 + v_s^3 \rho + \frac{3}{2}v_s^2 \rho^2 + \frac{1}{2}v_s^2 \phi^2.
\end{align}

\begin{align}
	S^2 + S^{\dagger 2} &= \frac{1}{2}(v_s + \rho + i\phi)^2 + \frac{1}{2}(v_s + \rho - i\phi)^2 \nonumber \\
	&= (v_s + \rho)^2 - \phi^2 \nonumber \\
	&= v_s^2 + 2v_s \rho + \rho^2 - \phi^2.
\end{align}

\subsection{The Potential at Quadratic Order}

Substituting these expansions back into $V_S$, we isolate the constant (vacuum), linear (tadpole), and quadratic (mass) pieces:
\begin{align}
	V_S &\supset \rho \left[ v_s \mu_S^2 + \frac{\lambda_S}{2}v_s^3 + v_s \mu_{sb}^2 \right] \nonumber \\
	&\quad + \frac{1}{2} \rho^2 \left[ \mu_S^2 + \frac{3}{2}\lambda_S v_s^2 + \mu_{sb}^2 \right] \nonumber \\
	&\quad + \frac{1}{2} \phi^2 \left[ \mu_S^2 + \frac{1}{2}\lambda_S v_s^2 - \mu_{sb}^2 \right].
\end{align}

\subsection{Vacuum Extremization and Tadpole Cancellation}

For the vacuum to be stable, the linear term in $\rho$ (the tadpole) must vanish:
\begin{equation}
	\frac{\partial V}{\partial \rho} = 0 \implies v_s \left(\mu_S^2 + \mu_{sb}^2 + \frac{\lambda_S}{2}v_s^2\right) = 0.
\end{equation}
Assuming $v_s \neq 0$, we obtain the critical tadpole relation:
\begin{equation}
	\mu_S^2 + \frac{\lambda_S}{2}v_s^2 = -\mu_{sb}^2.
	\label{eq:tadpole}
\end{equation}

\subsection{Final Mass Eigenvalues}

We now substitute the tadpole relation~(\ref{eq:tadpole}) into the quadratic coefficients to find the physical masses squared $m^2 = \partial^2 V / \partial \varphi^2$.

\paragraph*{Radial mode $\rho$:}
\begin{align}
	m_\rho^2 &= \mu_S^2 + \mu_{sb}^2 + \frac{3}{2}\lambda_S v_s^2 \nonumber \\
	&= \left( -\frac{\lambda_S}{2}v_s^2 - \mu_{sb}^2 \right) + \mu_{sb}^2 + \frac{3}{2}\lambda_S v_s^2 = \lambda_S v_s^2.
\end{align}

\paragraph*{pNGB mode $\phi$:}
\begin{align}
	m_\phi^2 &= \mu_S^2 - \mu_{sb}^2 + \frac{1}{2}\lambda_S v_s^2 \nonumber \\
	&= \left( -\frac{\lambda_S}{2}v_s^2 - \mu_{sb}^2 \right) - \mu_{sb}^2 + \frac{1}{2}\lambda_S v_s^2 \nonumber \\
	&= -2\mu_{sb}^2 = 2|\mu_{sb}^2|,
	\label{eq:app_phi_mass_final}
\end{align}
where the equivalence in the final step follows from the convention $\mu_{sb}^2 < 0$, which is required to ensure a positive mass-squared for the dark energy field ($m_\phi^2 > 0$). This completes the derivation.

\end{document}